\documentclass[a4paper,fleqn,usenatbib]{mnras}

\usepackage{newtxtext,newtxmath}

\usepackage[T1]{fontenc}
\usepackage{ae,aecompl}
\usepackage{booktabs}

\usepackage{graphicx}	
\usepackage{amsmath}	
\usepackage{amssymb}	
\usepackage{color}
\usepackage{xspace}
\usepackage{bm}
\usepackage[normalem]{ulem}
\usepackage{physics}

\newcommand{\kv}{\pmb{k}}	

\newcommand{\ko}{\pmb{k_{1}}}	
\newcommand{\kt}{\pmb{k_2}}
\newcommand{\kth}{\pmb{k_3}}
\newcommand{\sv}{\pmb{s}}
\newcommand{\rv}{\pmb{r}}
\newcommand{\V}{V_{\rm eff}}

\newcommand{\msun}{{\rm M}_{\odot}}

\title[Phase statistics of non-standard cosmology]{Cosmological Constraints from Fourier Phase Statistics}

\author[Ali et al.]{Kamran Ali$^{1,2}$,
Danail Obreschkow$^{1,2}$,
Cullan Howlett$^{1,2}$,
Camille Bonvin$^{3}$,
\newauthor
Claudio Llinares$^{4}$,
Felipe Oliveira Franco$^{3}$,
Chris Power$^{1,2}$
\\
\\
$^{1}$International Centre for Radio Astronomy Research (ICRAR), University of Western Australia, 35 Stirling Highway,\\
~Crawley WA 6009, Australia\\
$^{2}$ARC Centre of Excellence for All-sky Astrophysics (CAASTRO) \\
$^{3}$D\'{e}partment de Physique Th\'{e}orique and Center for Astroparticle Physics (CAP), University of Geneva, \\
~24 quai Ernest Ansermet, CH-1211 Geneva, Switzerland\\
$^{4}$Institute for Computational Cosmology, Department of Physics, Durham University, Durham DH1 3LE, U.K.
}

\date{Accepted XXX. Received YYY; in original form ZZZ}

\pubyear{2017}
\begin{document}
\label{firstpage}
\pagerange{\pageref{firstpage}--\pageref{lastpage}}
\maketitle

\begin{abstract}
Most statistical inference from cosmic large-scale structure relies on two-point statistics, i.e.\ on the galaxy-galaxy correlation function (2PCF) or the power spectrum. These statistics capture the full information encoded in the Fourier amplitudes of the galaxy density field but do not describe the Fourier phases of the field. Here, we quantify the information contained in the line correlation function (LCF), a three-point Fourier phase correlation function. Using cosmological simulations, we estimate the Fisher information (at redshift $z=0$) of the 2PCF, LCF and their combination, regarding the cosmological parameters of the standard $\Lambda$CDM model, as well as a Warm Dark Matter (WDM) model and the $f(R)$ and Symmetron modified gravity models. The galaxy bias is accounted for at the level of a linear bias. The relative information of the 2PCF and the LCF depends on the survey volume, sampling density (shot noise) and the bias uncertainty. For a volume of $1h^{-3}\rm Gpc^3$, sampled with points of mean density $\bar{n} = 2\times10^{-3} h^{3}\ \rm Mpc^{-3}$ and a bias uncertainty of 13\%, the LCF improves the parameter constraints by about 20\% in the $\Lambda$CDM cosmology and potentially even more in alternative models. Finally, since a linear bias only affects the Fourier amplitudes (2PCF), but not the phases (LCF), the combination of the 2PCF and the LCF can be used to break the degeneracy between the linear bias and $\sigma_8$, present in 2-point statistics.
\end{abstract}



\section{Introduction}

Cosmic large-scale structure (LSS) grows from primordial density fluctuations under the effect of gravity and dark energy. This structure hence contains useful information on the cosmological model, whether this is a specific form of the standard $\Lambda$CDM model or an alternative proposal. Modern galaxy redshift surveys \citep[e.g.][]{Alam2017MNRAS.470.2617A, Des2017arXiv170801530D} decode the information contained in the LSS using spatial statistics, most commonly the two-point statistics, that is the isotropic two-point correlation function (2PCF) or its Fourier counterpart, the power spectrum.

While the structural information of the early universe, seen in the cosmic microwave background (CMB), seems to be fully described by the two-point statistics, non-linear gravity-driven evolution causes a flow of information into higher order statistics as the universe evolves \citep{Scoccimarro1997ApJ...487....1S}. This motivates the search for efficient statistical estimators to probe the excess information in the LSS which lies beyond the two-point statistics. This quest has gained much momentum with the prospect of further surveys such as DES \citep{Des2005astroph10346T}, TAIPAN  \citep{taipan2017arXiv170601246D}, EUCLID \citep{EUCLID2011arXiv1110.3193L} and SKA HI surveys \citep{ska2009IEEEP..97.1482D}.

Obvious candidates for the statistical estimators beyond the 2PCF are the three-point and higher order isotropic correlation functions. In fact the family of all $N$-point correlation functions constitutes a full (albeit highly redundant) description of the statistical information in the LSS. For instance the bispectrum (Fourier transform of the three-point correlation function) of the CMB sets an upper limit on the non-Gaussianity \citep{2016A&A...594A..13P} of the primordial density field, thereby restricting the range of allowed cosmological models. At late times, the bispectrum provides additional constraints on the standard cosmological model \citep{Gil2015MNRAS.451..539G}. The same estimator, however, is unable to put stringent constraints on alternate cosmological models \citep{GilMarin2011JCAP...11..019G}, until the effects of redshift space distortions are incorporated into the analysis. The information pertaining to the redshifted structure allows differentiation between modified gravity and standard models \citep{Sabiu2016A&A...592A..38S} at small scales. Including higher order ($>3$ point) correlation functions makes it possible to increase the minimum length scales that need to be considered to differentiate between cosmological models \citep{Hellwing2013MNRAS.435.2806H}. \\

Alternative estimators to the standard $N$-point correlation functions exist and have previously been used to distinguish between standard and alternate cosmological models. For instance, halo shape statistics \citep{Llinares2014A&A...562A..78L}, the two-point function of different types of non-linearly rescaled density fields \citep{Llinares2017arXiv170402960L,White2016JCAP...11..057W,Lombriser2015PhRvL.114y1101L} and void count and shape statistics \citep{Voivodic2017PhRvD..95b4018V, Cai2015MNRAS.451.1036C, Falck2017arXiv170408942F} also probe this non-linear regime and allows us to differentiate between modified gravity models. However, these alternative estimators, as well as the standard $N$-point correlation functions, are strongly correlated to the 2PCF. This is because these alternative estimators depend on the amplitudes of the Fourier field, already fully measured by the 2PCF. Therefore, the total constraints from the 2PCF and such alternative estimators is often smaller than their independent addition would suggest.

Given the dependence of the 2PCF on the Fourier space amplitudes, it seems promising to introduce a second statistical measure that only depends on Fourier phases. One such measure is the so-called line correlation function (LCF, defined in Section~\ref{ss:estimators}) introduced by \citet{Obreschkow2013ApJ...762..115O}. Using a simplistic information analysis \citeauthor{Obreschkow2013ApJ...762..115O} speculated that the LCF is a promising estimator, especially when probing alternative cosmological models. Moreover, because the LCF measures the three-point statistics of the Fourier phases, irrespective of amplitudes, it is independent of linear bias \citep{Wolstenhulme2015ApJ...804..132W}, the uncertainty of which plagues all LSS surveys. To study the effectiveness of the LCF in galaxy surveys, \citet{Eggemeier2017MNRAS.466.2496E} studied its correlation with the 2-point estimator on different scales. This was further expanded upon by \citet{Byun2017MNRAS.471.1581B} who compared the effectiveness of the LCF with other 3-point estimators, including the bispectrum and found it to be a promising candidate for future surveys. Given that all the odd isotropic $N$-point functions contribute to the LCF \citep{Wolstenhulme2015ApJ...804..132W}, we expect this estimator to play a more significant role in cosmological models which affect the non-linear regime of structure evolution due to gravity. Hence, it is an interesting avenue to apply the LCF to alternative gravity models.

In this paper we investigate the additional information contained in the LCF, relative to the information in the 2PCF, based on cosmological $N$-body simulations. We account for the difference between observable galaxies and the underlying total density field via a linear bias. We start by defining the statistical estimators and the Fisher Information (FI) Matrix (FIM). We then set out the algorithm to compute the derivatives and covariances required for the FIM and investigate the effect of the uncertainty of the linear bias on the overall covariance matrix. In Section \ref{sec:sims} we use $N$-body simulations to measure the information in the 2PCF, the LCF and their combined information in standard ($\Lambda$CDM), Warm Dark Matter (WDM), $f(R)$ and Symmetron cosmologies and end by exploring the effect of linear bias uncertainties on parameter estimation. Throughout this work, we only take into account the real-space 2PCF and LCF while similar analyses in redshift-space are left for future work. Section~\ref{sec:conclusion} concludes the paper with a brief discussion.

\section{methods} \label{sec:method}

This section introduces the tools used for analysing the cosmological information in $N$-body simulations. We start by defining the statistical estimators, i.e.~the 2PCF and LCF, and outlining the method of measuring their relative information. We then describe the algorithm used to measure the derivatives and covariances of the 2PCF and LCF required for FIM. Finally, we quantify the effect of linear bias uncertainties on the covariance matrix of the statistical estimators.

\subsection{Estimators}\label{ss:estimators}


Cosmological $N$-body simulation boxes with periodic boundary conditions (see Section \ref{subsec:lcdm}) are used to compute the matter density field $\rho(\pmb{r})$ at redshift $z=0$. It is convenient to define the over-density field, $\delta(\pmb{r})$, as
\begin{equation} \label{eq:delta}
    \delta(\pmb{r}) = \frac{\rho(\pmb{r})-\bar{\rho}}{\bar{\rho}},
\end{equation}
where $\bar{\rho}$ is the average density within the simulation volume. The next step is to take the Fourier transform of this quantity, using the formalism of \citet{Obreschkow2013ApJ...762..115O}, and compute the power spectrum, $P(\kv)$, defined as
\begin{equation} \label{eq:pk}
    \left\langle\delta_{\pmb{k}}\delta_{\pmb{k^{\prime}}}\right\rangle =(2\uppi)^{3}\updelta_{\rm D}(\pmb{k}+\pmb{k^{\prime}})P(\pmb{k});
\end{equation}
where $\langle\rangle$ and $\updelta_{\rm D}$ are the ensemble averages and Dirac Delta function, respectively. The 2PCF is then obtained using
\begin{equation} \label{eq:2pcf}
    \xi(r) = \frac{V}{(2\uppi)^{3}} \int d^3k \frac{\sin(|\kv|r)}{|\kv|r} P(|\kv|).
\end{equation}
Since the 2PCF and the power spectrum are Fourier counterparts of one another they contain the same FI. Note that in real surveys this might not be the case given that the two statistics are computed using their own approximate estimators.

The LCF measures correlation between the Fourier phases, $\epsilon(\pmb{k})= \delta(\pmb{k})/|\delta(\pmb{k})|$. We here use the LCF definition of \citet{Wolstenhulme2015ApJ...804..132W}
\begin{equation} \label{eq:lcf}
 \begin{split}
    \ell(r) & = \frac{V^{3}}{(2\uppi)^9} \left(\frac{r^3}{V}\right)^{3/2} \iint\limits_{|\ko|,|\kt|,|\kth|\leq2\uppi/r} d^3\ko d^3\kt d^3\kth \\ 
    & e^{i\left[\ko\cdot\sv+\kt\cdot(\sv+\rv)+\kth\cdot(\sv-\rv)\right]}\left\langle\epsilon(\ko)\epsilon(\kt)\epsilon(\kth)\right\rangle.
  \end{split}
\end{equation}
This LCF is computed using the \texttt{ProCorr} package, which follows the formalism of \citet{Obreschkow2013ApJ...762..115O}.

\subsection{Fisher Information Matrix}
\label{sec:fisher} 

The main aim of this paper is to measure the available information in the estimators and thereby constrain a set of parameters. To do so we use the FIM, which for a given Log-Likelihood function, $\mathcal{L}$, is defined by 
\begin{equation} \label{eq:genfisher}
    \mathbf{F}_{\rm ij}=\left\langle \frac{\partial \mathcal{L}(\bm{P}, \bm\theta)}{\partial \theta_{\rm i}} \frac{\partial \mathcal{L}(\bm{P}, \bm\theta)}{\partial \theta_{\rm j}} \right\rangle,
\end{equation}
where $\mathcal{\bm\theta}$ consists of $N$ model parameters, $\theta_{\rm i} \forall\ i\in1,...,N$, with $\bm{P}$ being the estimator we measure. A useful feature of this matrix is that its inverse is an estimator of the covariance matrix of the model parameters (see Section~\ref{subsec:lcdm}). In the Laplace approximation (Gaussianity of likelihoods), the FIM simplifies to \citep{Tegmark1997ApJ...480...22T}
\begin{equation} \label{eq:fisher}
\begin{split}
    \mathbf{F}_{\rm ij} & =  \frac{\partial \bm{P}(\mathcal{\bm\theta})}{\partial \theta_{\rm i}} \mathbf{C}(\bm{P,\bm\theta})^{-1}\frac{\partial \bm{P}(\mathcal{\bm\theta})}{\partial \theta_{\rm j}} + \\
    & \frac{1}{2} {\rm Tr} \left[ \mathbf{C}(\bm{P,\theta})^{-1} \frac{\partial \mathbf{C}(\bm{P,\bm\theta})}{\partial\theta_{\rm i}} \mathbf{C}(\bm{P,\bm\theta})^{-1} \frac{\partial \mathbf{C}(\bm{P,\bm\theta})}{\partial\theta_{\rm j}} \right],
\end{split}
\end{equation}
where $\mathbf{C}$ is the covariance matrix of the likelihood. The variation of the covariance matrix (trace term in equation~\ref{eq:fisher}) turns out to be subdominant relative to the first term. In the case of the 2PCF, this was shown by explicit numerical calculations. For instance, in the particular case of the Symmetron modified gravity model, \cite{Llinares2017arXiv170402960L} found that the covariance matrix of the power spectrum (hence the 2PCF) does not vary significantly relative to that of $\Lambda$CDM. For the LCF, \cite{Eggemeier2017MNRAS.466.2496E} show that the trace term vanishes identically at lowest order, since the Gaussian part of the covariance matrix is independent of the cosmology. This lowest order solution agrees with full numerical computations over a wide range of length scales $r\geq 40\ h^{-1}\rm Mpc$. Hence, in subsequent sections, we will ignore the trace term and evaluate the derivatives of the 2PCF and the LCF relative to the model parameters at a fiducial cosmology.

An alternative information measure, commonly, used in cosmology literature is the cumulative signal-to-noise of the relevant statistic \citep[e.g.][]{Bonvin2016JCAP...08..021B, Sefusatti2011JCAP...03..047S}. This metric, defined in terms of the raw estimator instead of its derivative, i.e.~${\rm SNR}=\bm{P}\mathbf{C}^{-1}\bm{P}$, quantifies the possibility of measuring the estimator in a given cosmological volume. However, in general the Fisher methodology provides a more robust way of estimating the parameters.

\subsection{Derivatives} \label{subsec:derivatives}

To evaluate the FI of the LCF and the 2PCF we require the expectation values of their derivatives with respect to the cosmological parameters $\theta_i$ under investigation. In practice these expectations of the derivatives can be computed with a finite difference method using two simulation boxes per parameter, one centred at a fiducial cosmology, $\bm\theta_{\rm 0}$, and one with the parameter in question varied by a small difference $\Delta\theta$,
\begin{equation} \label{eq:derivative}
    \frac{\partial \bm{P}(\theta_0)}{\partial \theta} =  \frac{\bm{P}(\theta_0)-\bm{P}(\theta_0-\Delta\theta)}{\Delta\theta}.
\end{equation}
Here, the two simulations (with $\theta_0$ and one for $\theta_0-\Delta\theta$) rely on the same initial random seed to minimise the randomness of the derivatives. We further reduce this randomness by applying a slight smoothing filter (2-point running average) to the LCF, which suffers from the limited number of modes more than the 2PCF. In section \ref{ss:linbias}, where we fit the bias in $\Lambda$CDM only at large scales (with only few modes per box), the derivatives are averaged over 50 random realisations.

\subsection{Covariance matrix}\label{ss:cov}

\begin{figure}
	\includegraphics[width=\columnwidth]{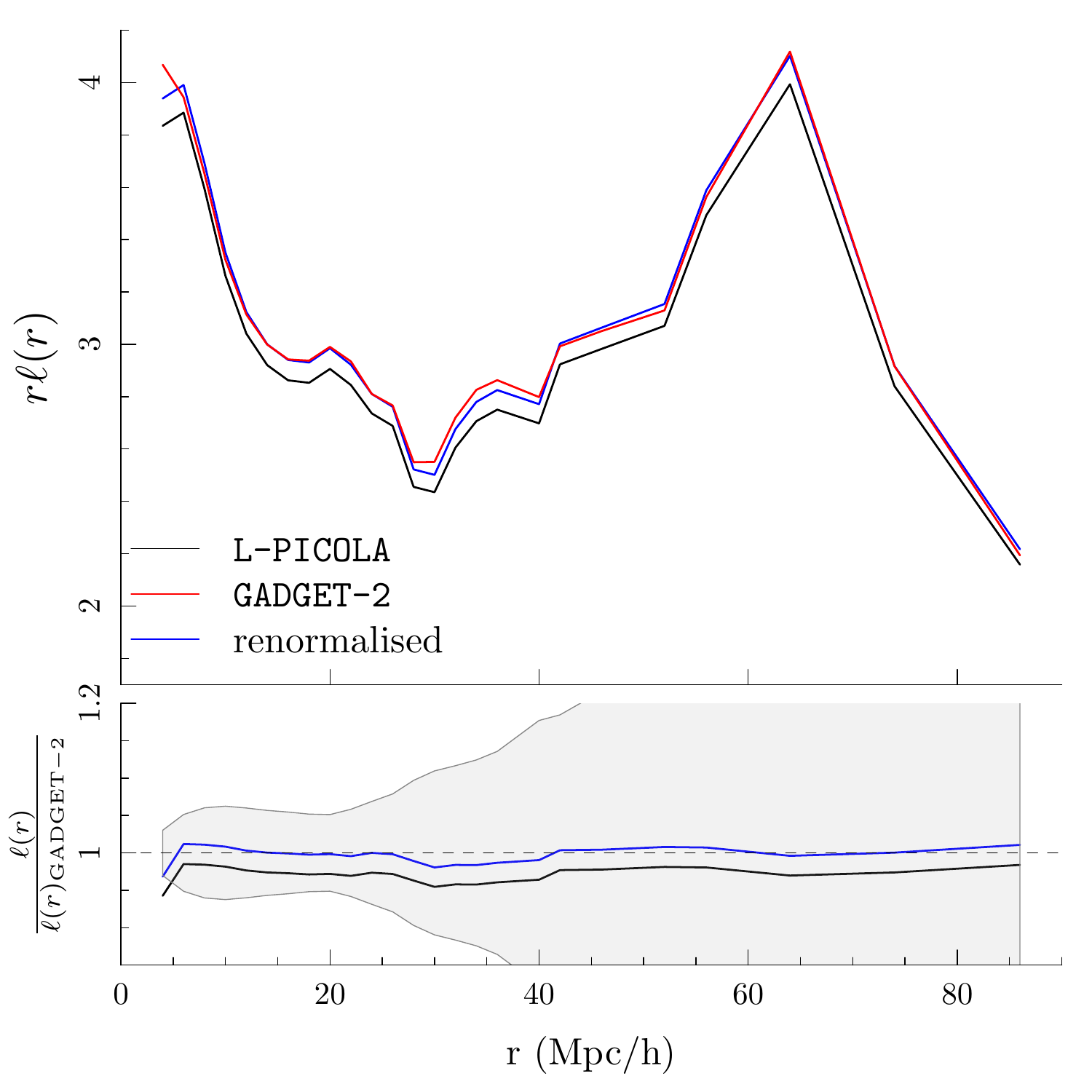}
    \caption{Comparison of the LCF of simulation boxes evolved, by the two $N$-body codes \texttt{L-PICOLA} and \texttt{GADGET-2}, from the same initial density field. \texttt{L-PICOLA} tends to slightly but systematically underestimate the LCF by $\approx 3\%$ which needs to be accounted for before computing the covariance matrix. The shaded region shows the standard deviation of LCF indicating the significance of the systematic correction at different scales.} 
    \label{fig:lr_accuracy}
\end{figure}

\begin{figure*}
	\includegraphics[trim={0 1.659cm 0 2.259cm},clip,scale=0.85]{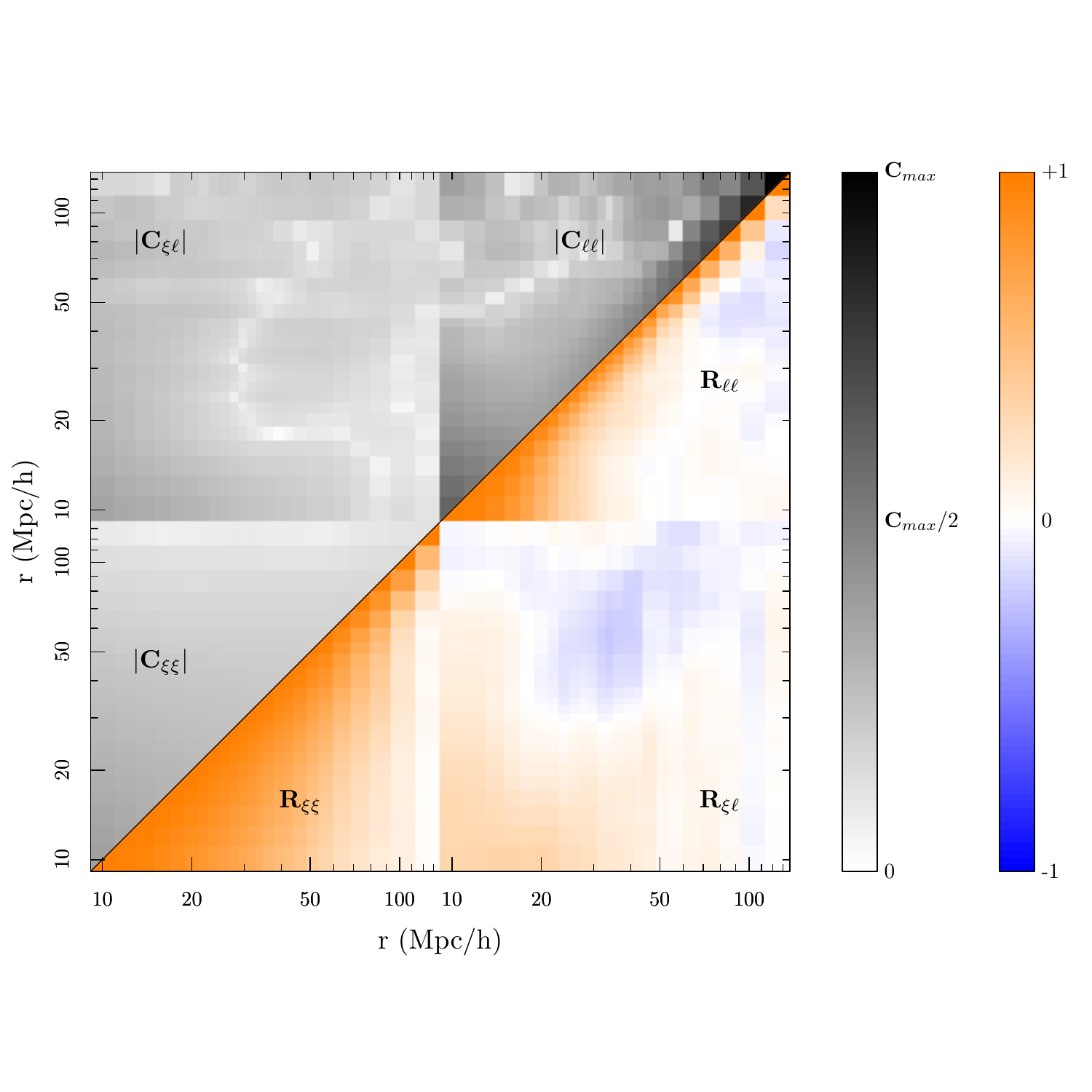}
    \caption{The absolute value of covariance matrix (upper triangle) estimated from 500 \texttt{L-PICOLA} simulations and its associated correlation matrix (lower triangle) computed by inverting equation~(\ref{eq:corrmat}) as a function of scale length. As we go to larger scales the variance of the LCF increases while that of 2PCF decreases. The correlation matrix, on the other hand, increases at small scales for all combination of the parameters 2PCF-2PCF (bottom-left), 2PCF-LCF (bottom-right) and LCF-LCF (top-right). The increase in 2PCF-LCF correlation at $r\lesssim 20 h^{-1}\rm Mpc$ indicates strong coupling between Fourier phases and amplitudes. This property is absent in the early universe and results from non-linear growth of structure due to gravity.}
    \label{fig:covmat}
\end{figure*}

The second ingredient for the FIM computation is the covariance matrix of the respective statistical estimator. We use 500 simulation boxes which are generated using the COmoving Lagrangian Acceleration (COLA) method \citep{Tassev2013JCAP...06..036T} implemented in the \texttt{L-PICOLA} code \citep{Howlett2015A&C....12..109H}. The simulations consist of $256^3$ particles enclosed in a periodic box of length $L=512\ h^{-1}\rm Mpc$ with the same background cosmology as in \citet{Howlett2015A&C....12..109H}. The LSS evolved by the COLA technique provides an accurate estimate of the covariances of two-point statistics to scales of $k \leq 0.3\ h\ \rm{Mpc}^{-1}$. Hence, in our regime of interest, $r \geq 10 \ h^{-1}\rm{Mpc}$, we can safely use \texttt{L-PICOLA} to estimate the covariance matrix of the 2PCF. 

The three-point statistics on the other hand depend strongly on the non-linear growth of gravity and, hence, might not be easily reproducible by the COLA code. To gauge the accuracy of the COLA solver we compare a set of simulation boxes which are evolved from the same initial density field but by two different $N$-body codes, \texttt{GADGET-2} and \texttt{L-PICOLA}. As shown in Fig.~\ref{fig:lr_accuracy}, \texttt{L-PICOLA} slightly, but systematically underestimates the LCF by about $3\%$. We approximately correct this small inaccuracy by rescaling the \texttt{L-PICOLA} LCF by $3\%$ in computing the covariance matrix.  This correction is more significant at smaller scales, but the difference persists even at larger scales and might be due to \texttt{L-PICOLA} being unable to accurately reproduce the non-linear regime of structure growth since the LCF depends entirely on the correlation between Fourier phases, the growth of which is a purely non-linear phenomenon. We here assume that the applied correction also holds for the covariance matrix estimated through the \texttt{L-PICOLA} simulations. Since, the LCF depends on three-point statistics, the resulting covariance matrix should depend on four, five and six-point functions which might differ by a different factor in the COLA method. We expect these corrections to be small relative to the diagonal terms in the covariance matrix; hence they do not significantly alter the FIM.

Given 500 \texttt{L-PICOLA} realisations of the 2PCF and the LCFs, we estimate the full covariance matrix
\begin{equation} \label{eq:covmat}
     \mathbf{\hat{C}} = \begin{bmatrix}
    				\mathbf{\hat{C}_{\rm \xi\xi}} & \mathbf{\hat{C}_{\rm \xi\ell}}  \\
    				\mathbf{\hat{C}_{\rm \ell\xi}} & \mathbf{\hat{C}_{\rm \ell\ell}}
\end{bmatrix},
\end{equation}
where the four sub-matrices are defined by
\begin{equation}
\begin{split}
     (\mathbf{\hat{C}_{\rm \xi\xi}})_{\rm ij} &= {\rm cov}_{\rm N_{\rm sim}} \left(\xi(r_{\rm i}),\xi(r_{\rm j})\right) \\
     (\mathbf{\hat{C}_{\rm \ell\ell}})_{\rm ij} &= {\rm cov}_{\rm N_{\rm sim}} \left(\ell(r_{\rm i}),\ell(r_{\rm j})\right) \\
     (\mathbf{\hat{C}_{\rm \xi\ell}})_{\rm ij} &= (\mathbf{\hat{C}_{\rm \ell\xi}})_{\rm ji} = {\rm cov}_{\rm N_{\rm sim}} \left(\xi(r_{\rm i}),\ell(r_{\rm j})\right).
     \end{split}
\end{equation}
The operator ${\rm cov}_{\rm N_{\rm sim}}$ is the estimator of the covariance matrix from $N_{\rm sim}$ simulation boxes,
\begin{equation}\label{eq:cov}
{\rm cov}_{\rm N_{\rm sim}}(x,y) = \frac{1}{N_{\rm sim}-1} \sum_{ k=1}^{N_{\rm sim}} \left(x_{\rm k}-\langle x\rangle\right)\left(y_{\rm k}-\langle y\rangle\right).
\end{equation}

The covariance matrix of cosmological surveys is, in general, made up of three terms (see \cite{Bonvin2016JCAP...08..021B}): a Poisson (or shot noise) contribution which depends on the density of the observed galaxies. The second major contribution comes from the fact that we observe a single realisation of the observable universe. The final contribution to the overall covariance matrix is a mixture of the Poisson and cosmic effect and hence depends on both the galaxy density and the survey volume. In the absence of a window function, the covariance matrices of the 2PCF \citep{Meiksin:1998mu, Scoccimarro1999ApJ...527....1S, Howlett2017MNRAS.472.4935H} and the LCF \citep{Eggemeier2017MNRAS.466.2496E} are inversely proportional to the survey volume. 
This allows us to assume that the covariance matrices of the 2PCF and LCF both scale as the inverse of the survey volume, i.e.\ the covariance matrix $\mathbf{\tilde{C}}$ for a survey of volume $\V$ is
\begin{equation} \label{eq:Vfactor}
\mathbf{\tilde{C}} = \frac{V_{\rm sim}}{\V} \mathbf{\hat{C}},
\end{equation}
where $V_{\rm sim}$ is the simulation volume in which the raw covariance matrix $\mathbf{\hat{C}}$ was evaluated.

To gauge the effect of shot noise in our analysis we randomly draw a subset of $N_{\rm parts}$ particles from the simulation box, such that the mean expected particle density $\bar{n} = N_{\rm parts}/V_{\rm sim}$ equals $2\times10^{-3} h^{3}\ \rm Mpc^{-3}$, approximately corresponding to the cumulative space density of haloes more massive than $10^{12}\msun$ \citep{Murray2013A&C.....3...23M} -- the dynamical mass of typical $M^\ast$-galaxies. Strictly speaking, this subsampling process mimics shot noise in the matter field rather than in the galaxy field -- a difference, which we neglect in this work, similarly to others \citep{Eggemeier2017MNRAS.466.2496E}. By subsampling the simulation box in this way, we find that the variance of the 2PCF estimator increases while that of the LCF decreases. This is because the expectation of the 2PCF is invariant to random subsampling of the particle field, while that of the LCF decreases significantly, a behaviour explored by \citet{Eggemeier2017MNRAS.466.2496E}. Note that in computing the FIM in the presence of shot noise, we rescale the derivatives of the full LCF (without shot noise) by a factor $f(r)$, defined as the ratio between the expectation of the LCF of the sub-sampled boxes and the full LCF. This approach neglects the derivative of $f(r)$ with respect to the cosmological parameters. Numerically, this simplification has a relatively small ($\lesssim20\%$) effect on the overall FI in the LCF. (An exact computation would nonetheless require more simulation boxes than available for this analysis.) 

Modern galaxy redshift surveys probe cosmic volumes of a few $h^{-3} \rm Gpc^3$. For instance, the CMASS galaxy sample of the BOSS \citep{BOSS2011AJ....142...72E} survey covers an effective volume of $3\ h^{-3}{\rm Gpc}^3$ \citep{Ntelis2017JCAP...06..019N} while future surveys might reach up to $20\ h^{-3} \rm Gpc^3$ \citep{Duffy2014AnP...526..283D}. In this work all results are presented for a reference volume of $\V = 1\ h^{-3}{\rm Gpc}^3$.\\

While the covariance matrix computed via equation~(\ref{eq:cov}) is an unbiased estimator, its inverse required for the FIM is generally not. To obtain an unbiased estimator, we apply the correction of \cite{2007A&A...464..399H}
\begin{equation} \label{eq:ahfactor}
    \mathbf{C}^{-1} = \frac{N_{\rm sim} - N_{\rm dim} -2}{N_{\rm sim}-1}\  \mathbf{\tilde{C}}^{-1},
\end{equation}
where $N_{\rm dim}$ is the order of $\mathbf{\tilde C}$.

Fig.~\ref{fig:covmat} shows the covariance matrix $\mathbf{C}$ (upper triangle), as well as its correlation matrix $\mathbf{R}$ (lower triangle), defined as
\begin{equation} \label{eq:corrmat}
   \mathbf{R} = \mathbf{D}^{-1}\mathbf{C}\mathbf{D}^{-1},
\end{equation}
where $\mathbf{D}$ is the square root of the diagonal matrix of $\mathbf{C}$. 

Fig.~\ref{fig:covmat} reveals that the cross-correlation between the 2PCF and the LCF is small, as expected from the fact that the former estimator measures the Fourier amplitudes while the latter measures the Fourier phases. Only at scales $\lesssim 20\ h^{-1}\rm Mpc$ does the cross-correlation become significant, indicating a strong coupling between the Fourier phases and amplitudes due to non-linear growth. Note that the rescaling of equations~\ref{eq:Vfactor} and \ref{eq:ahfactor} has no effect on the correlation matrix. 

\subsection{Incorporating bias uncertainty} \label{subsec:bias_definition}

Most cosmological surveys use galaxies as tracers of the matter density field $\delta(\pmb{r})$. The mapping between the galaxy density $\delta_g(\pmb{r})$ and $\delta(\pmb{r})$ is complicated in detail. However, at large scales, the asymptotic effect is a uniform rescaling 
\begin{equation}
    \delta_g(\pmb{r}) = b\ \delta(\pmb{r}),
\end{equation}
where the scaling factor $b$ is known as linear bias. This bias only affects the amplitudes, not the phases of the Fourier modes, i.e.\ $\delta_g(\mathbf{k})=b\delta(\mathbf{k})$. Therefore only the 2PCF is affected by linear bias, not the LCF. Explicitly, the galaxy 2PCF at large scales ($\gtrsim50h^{-1}Mpc$) is
\begin{equation} \label{eq:gal_xi2}
    \xi_{\rm g,i} = b^2\ \xi_i
\end{equation}
where $\xi_{\rm i} = \xi (r_{\rm i})$ is the estimated dark matter 2PCF at separation scale $r_{\rm i}$. On smaller scales, the bias becomes non-linear. The effects of this non-linearity on the LCF are not yet well understood and will be neglected in this work, although they might be worth considering in the future.

The FI of the 2PCF does not depend on the absolute value of $b$, since the covariance matrix $\mathbf{C}$ scales as $b^4$, while each of the two derivative terms scales as $b^2$, hence cancelling $b$ in equation~(\ref{eq:fisher}). Only the \emph{uncertainty} of $b$ affects the constraints on the cosmological parameters. We can account for this effect of linear bias uncertainty in two ways that correspond to slightly different concepts. The first approach consists of inferring the cosmological parameters from the \textit{galaxy field}, which depends both on these parameters and $b$. One then computes the combined FIM of the model parameters and $b$, hence constraining them simultaneously. The subtle caveat of this approach is that a linear bias $b$ only applies to the largest scales and hence should not be fitted to smaller scales. Ignoring this would result in significantly inflated constraints on $b$, although this problem can be somewhat alleviated by including a non-linear bias term \citep{Eggemeier2017MNRAS.466.2496E}. The alternative approach, used in this paper, is to assume that we infer the cosmological parameters from the \textit{dark matter field}, which is itself inferred by applying an uncertain bias to the galaxy field. This translates into an uncertain dark matter 2PCF. In this case the FIM is computed only for the cosmological parameters, but the bias uncertainty must be included in the estimator covariance matrix $\mathbf{C}$. This approach has the advantage that we can impose a realistic uncertainty for the linear bias, accounting for the complexity of real surveys. The downside is that this approach doesn't guarantee that the assumed bias uncertainty is consistent with the best estimate of the bias that one might achieve from combining the 2PCF and LCF at the largest scales. This consistency therefore needs to be checked in a separate step, see Section \ref{ss:linbias}.

Since the FIM does not depend on the absolute value of the linear bias, we can assume, without loss of generality, that $\langle b^2 \rangle=1$ and write the variance of $b^2$ as $\sigma^2_{\rm b^2}$. Using linear error propagation, the variance of the 2PCF with bias uncertainty then depends on $\sigma_{\rm b^2}$ via
\begin{equation} \label{eq:diagbias}
   \sigma^2_{\rm \xi_{\rm i}}(\sigma_{\rm b^2}) = \sigma^2_{\rm \xi_{\rm i}} + \sigma^2_{\rm b^2} \left( \xi^2_{\rm i}+ \sigma^2_{\rm \xi_{\rm i}}\right),
\end{equation}
where $\sigma^2_{\rm \xi_{\rm i}}$ are the diagonal elements of $\mathbf{C}_{\rm \xi\xi}$ without bias uncertainty. Note that this definition of $\sigma^2_{\rm \xi_{\rm i}}(\sigma_{\rm b^2})$ satisfies the condition $\sigma^2_{\rm \xi_{\rm i}}(0)=\sigma^2_{\rm \xi_{\rm i}}$. 

Equation~(\ref{eq:diagbias}) allows us to correct the diagonal elements of the covariance matrix $\mathbf{C}$ in order to account for bias uncertainty. The remaining question is how to correct the off-diagonal elements. We found that linear error propagation as in equation~(\ref{eq:diagbias}) is not the right approach. This is because the bias uncertainty introduces cross-correlation between different scales $r_{\rm i}$ and $r_{\rm j}$, which is not present in the estimations of the derivatives, when based only on \emph{one} random realisation of the universe (which necessarily can only have one bias value). To bypass this issue, we instead assume that the off-diagonal elements of the covariance matrix remain unchanged under the effect of bias uncertainties. Formally, the bias-corrected covariance matrix becomes

\begin{equation} \label{eq:corrmat2}
   \mathbf{C} (\sigma_{\rm b^2}) = \mathbf{D}(\sigma_{\rm b^2})\mathbf{R}\mathbf{D}(\sigma_{\rm b^2}),
\end{equation}

where $\mathbf{R}$ is the correlation matrix of $\mathbf{C}$ given in equation~(\ref{eq:corrmat}) and $\mathbf{D}(\sigma_{\rm b^2})$ is a diagonal matrix defined as

\begin{equation} \label{eq:diagonalscaling}
     \mathbf{D}(\sigma_{\rm b^2}) = \begin{bmatrix}
    				\mathbf{D}_{\rm \xi_g} (\sigma_{\rm b^2}) & \mathbf{0}  \\
    				\mathbf{0} & \mathbf{D}_{\rm \ell}
\end{bmatrix}
\end{equation}
with the diagonal sub-matrices
\begin{equation}
\begin{split}
     (\mathbf{D}_{\rm \xi_g})_{\rm ii} (\sigma_{\rm b^2}) &=  \sigma_{\rm \xi_{\rm g,i}} (\sigma_{\rm b^2}) \\
     (\mathbf{D}_{\rm \ell})_{\rm ii} &=  \sigma_{\rm \ell_{\rm i}}.
\end{split}
\end{equation}

Note that this definition of the covariance matrix $\mathbf{C} (\sigma_{\rm b^2})$ satisfies the condition $\mathbf{C} (0)=\mathbf{C}$, i.e.\ the original covariance matrix is recovered if the bias uncertainty vanishes. A consequence of this definition is the dependence of bias uncertainty on the galaxy survey volume as $1/\sqrt{V}$. We will use a fiducial survey volume throughout Section \ref{sec:sims} and in Section \ref{ss:linbias}  quantify the change in combined information of the estimators due to the bias uncertainty parameter.

\section{Application to Simulations} \label{sec:sims}

We will now evaluate the FIM of the 2PCF and LCF in different cosmological models using $N$-body simulations. All FIM computations are carried out at redshift $z=0$ and assume a fiducial survey volume of $\V = 1\ h^{-3}\rm Gpc^3$ and a linear bias uncertainty of 13\% (standard deviation), that is $\sigma_{\rm b^2}\approx0.26$. The results can be rescaled to other volumes and bias uncertainties using the covariance scaling relations given in Section \ref{subsec:bias_definition}. The smallest length scale accounted for in the statistical analyses is set at $r = 10\ h^{-1}\rm Mpc$. This is a conservative estimate that avoids most effects of (not modelled) non-gravitational baryonic physics, e.g.\ hydrodynamics and radiation.

\subsection{Standard $\Lambda$CDM cosmology} \label{subsec:lcdm}

\begin{figure*}
	\includegraphics[scale=0.8]{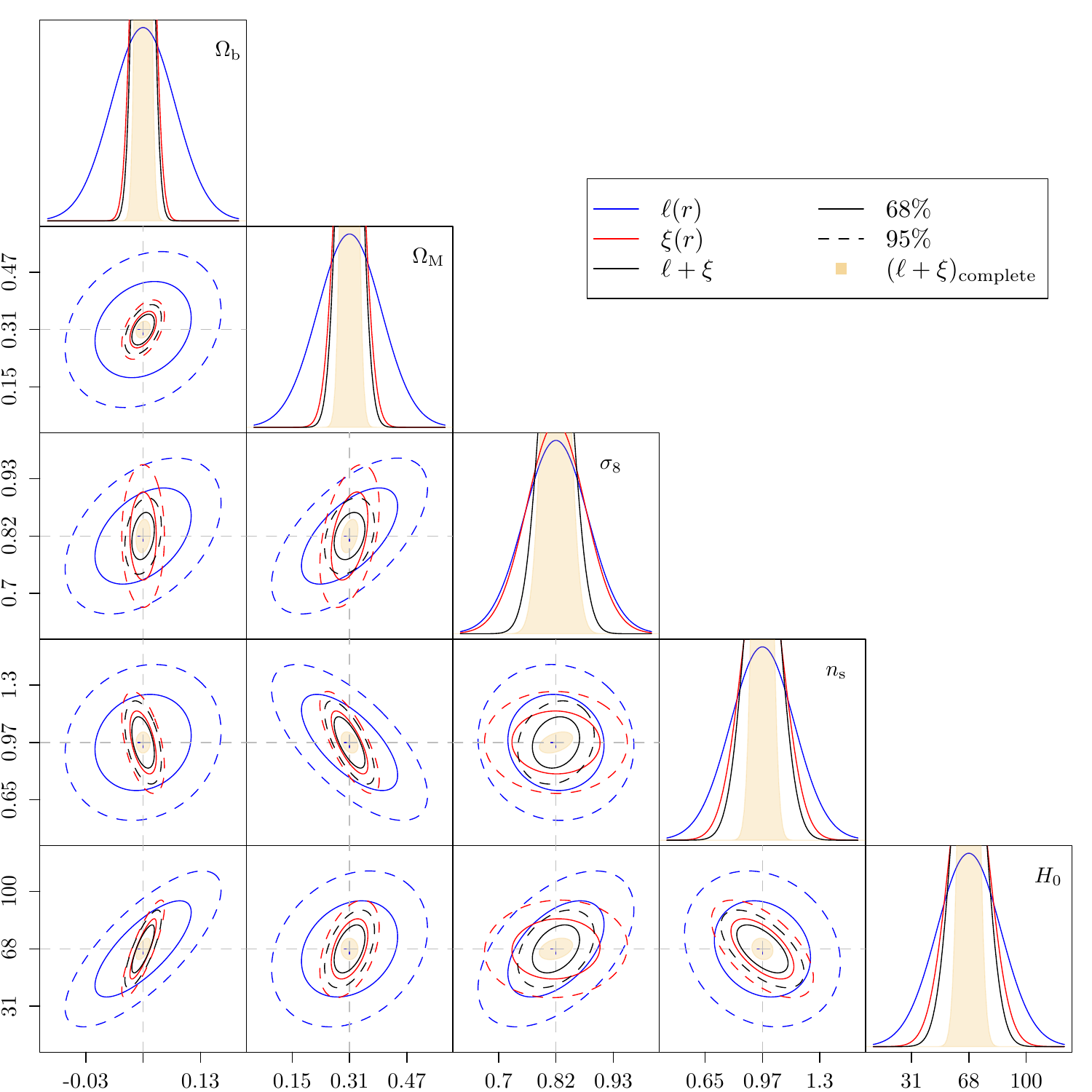}
	
    \caption{The constraints provided by the estimated 2PCF (red), LCF (blue) and their combination (black) for the standard cosmology parameters is shown on the left. The diagonal elements are shown as a 1-D Gaussian while the off-diagonal elements are depicted by their 68\% (solid) and 95\% (dashed) contours. The shaded yellow regions indicate combined constraints of the two estimators when all particles are used in the analysis (i.e.\ no shot noise) and as expected provides better constraints than the combined information from smaller particle density. A significant portion of the LCF ellipses have different orientations to those of the 2PCF allowing better constraints on the $\Lambda$CDM parameters. The highest gain in constraining power is placed on $\sigma_8$ with a \emph{gain} of $1.87$. For this analysis we assume a fiducial cosmological survey with $\sigma_{\rm b}=0.13$, $\bar{n}=2\times 10^{-3} h^{3} {\rm Gpc}^{-3}$ and $\V=1 h^{-3} {\rm Gpc}^3$  }
    \label{fig:lcdm_fi}
\end{figure*}

The bulk of the cosmological $N$-body simulations used to compute the estimator derivatives use $256^3$ particles in a periodic simulation box of side length $L=512\ h^{-1}\rm Mpc$. The particles are initially placed on a regular cartesian grid, then displaced to obey the power spectrum generated by \texttt{CAMB} \citep{Lewis2002PhRvD..66j3511L} and evolved using second order Lagrangian perturbation theory \citep{Crocce2006PhRvD..73f3519C} to redshift $z=49$. We then employ the \texttt{GADGET-2} \citep{Springel2005MNRAS.364.1105S} $N$-body solver to evolve the particles to redshift $z=0$.

To estimate the derivatives of the $\Lambda$CDM model we use equation~(\ref{eq:derivative}) with a background cosmology given by \citet{2016A&A...594A..13P} following which we rescale the LCF derivatives. We take the central parameter values of $\bm\theta_0 = \{ \Omega_{\rm b}, \Omega_{\rm M}, \sigma_{8}, n_{\rm s}, H_{0} ({\rm km/s}\ \rm Mpc^{-1}) \} = \{0.048, 0.31, 0.82, 0.97, 68 \}$ with a step spacing of $\Delta\bm\theta = \{0.10, 0.62, 0.86, 0.40, 46 \}\times 10^{-2}$. We do not include the reionisation depth parameter in our analysis since the linear matter power spectrum is independent of this parameter and thus it cannot be constrained by our methodology.

We then evaluate the FIM using the covariance matrix from the 500 \texttt{L-PICOLA} simulations (Sections~\ref{ss:cov} and \ref{subsec:bias_definition}). The covariance matrix of the model parameters, inferred from the statistical measurement, is then given by
\begin{equation}
	\bm\Sigma = \mathbf{F}^{-1}.
\end{equation}
The order of the matrix $\bm\Sigma$ is equal to the number of model parameters. The FIM can be computed using the 2PCF, the LCF or both of them simultaneously. To distinguish between the different parameter covariances resulting from these three cases, we use the symbols $\bm\Sigma_\xi$, $\bm\Sigma_\ell$, $\bm\Sigma_{\rm \xi\ell}$.

Fig.~\ref{fig:lcdm_fi} shows the parameter uncertainties implied by this covariance matrix in the Gaussian approximation, i.e.\ for a probability distribution $p(\bm\theta)\propto\exp[-(\bm\theta-\bm\theta_0)^\dag\bm\Sigma^{-1}(\bm\theta-\bm\theta_0)/2]$. The different line colors respectively show the constraints from the 2PCF, LCF and their combination. Interestingly, the uncertainty ellipses of the 2PCF and the LCF often have significantly different orientations. For instance in the case of the $\sigma_8$-$\Omega_{\rm M}$ pair, the LCF helps breaking the classic degeneracy. 

Generally a parameter $\theta_i$ is better constrained by the 2PCF+LCF than by the 2PCF alone. We quantify this \emph{gain} in constraining power as $g_{\rm i}=\sqrt{\bm\Sigma_{\rm \xi,ii}/\bm\Sigma_{\rm \xi\ell,ii}}$. For $\Lambda$CDM, we find that sub-sampling particles (i.e.\ mimicking shot noise) within the simulation box increases the contribution of the LCF to the combined constraints, $\bm{g}=\{1.17,1.19,1.87,1.23,1.26\}$ (for parameters $\Omega_{\rm b}, \Omega_{\rm M}, \sigma_{8}, n_{\rm s}, H_{0}$), compared to using the complete set of particles (i.e.\ no shot noise) in the analysis $\bm{g}_{\rm complete}=\{1.04,1.06,1.52,1.13,1.13\}$ for the same bias uncertainty. The same is not case when we neglect bias in the analysis since higher order functions are more susceptible to Poisson noise. This emphasises the importance of bias uncertainty in our analysis and we verify our choice of method in Section \ref{ss:linbias}.

Comparing our results with a previous study by \cite{Eggemeier2017MNRAS.466.2496E} we find a good match between the gain values of $\sigma_{8}$ and total matter density, $\Omega_{\rm M}$ (with shot noise), i.e.~$\bm{g}=\{1.25,1.90\}$ (for parameters $\Omega_{\rm M}, \sigma_{8}$). However, their gains on other parameters $\bm{g}=\{1.02,1.07, 1.07\}$ (for $\Omega_{\rm b}, n_{\rm s}, H_{0}$) are significantly smaller than those found by our full analysis. This disagreement might, primarily, be due to their study using CMB priors which put sharp constraints on $\Omega_{\rm b}$, $n_{\rm s}$ and $H_{0}$ parameters as compared to $\Omega_{\rm M}$ and $\sigma_{8}$. Furthermore, they incorporate the linear and non-linear bias into the FIM analysis and use $N$-body simulations at multiple redshifts to determine the overall parameter constraints.

Fig.~\ref{fig:cosmo_info_scale} (left) depicts the increase in the parameter uncertainty (decrease in information) as we increase the smallest length scale $r$ used in the FIM computation. One finding from this representation is that the relative constraining power from the LCF slowly increases with decreasing $r$ and for certain parameters contributes significantly to the combined constraints. For instance, the LCF provides similar constraints to the 2PCF for the $\sigma_8$ parameter while contributing at most half as much to the combined constraints of the other standard parameters. For all parameters, the increase in constraining power of the LCF at small scales is much steeper than that of the 2PCF due to the onset of non-linear regime.

\subsection{WDM cosmology} \label{subsec:wdm}

As a first alternative cosmological model, we consider the WDM model, often evoked as a possible solution to potential sub-structure problems in CDM \citep{Klypin1999ApJ...522...82K}.

Our approach to modelling a WDM universe consists in reducing the power of the initial CDM density field on small scales to mimic the free-streaming of the WDM. This is achieved by truncating the input linear matter power at large $k$ values (small scales), following  \cite{Bode2001ApJ...556...93B},

\begin{equation}\label{eqn:wdm}
    P(k)_{\rm WDM} =  \left[   1 + \left(\alpha k \right)^{2\nu}  \right]^{-5\nu} \ P(k)_{\rm CDM},
\end{equation}
where $P(k)$ is the linear CDM power spectrum, $\nu$ is a numerical constant and $\alpha$ is a non-linear function of the dark matter particle mass such that $\lim_{m_{\rm WDM} \to \infty} \alpha = 0$ and we recover the standard cosmology. We use the numerically fitted value of  $\nu=0.5$ by \cite{Bode2001ApJ...556...93B}. In the FI analysis, we consider $m^{-1}_{\rm WDM}$ as the free additional cosmological parameter to be constrained. To compute the estimator derivatives we use a spacing of $\Delta m^{-1}_{\rm WDM} =(0.2\ \rm keV)^{-1}$ centred at the a cosmology with $m^{-1}_{\rm WDM}=(0.2\ \rm keV)^{-1}$ and the same $\Lambda$CDM parameters as in the previous section. 

\begin{figure*}
\begin{tabular}{c c}
	\includegraphics[scale=0.525]{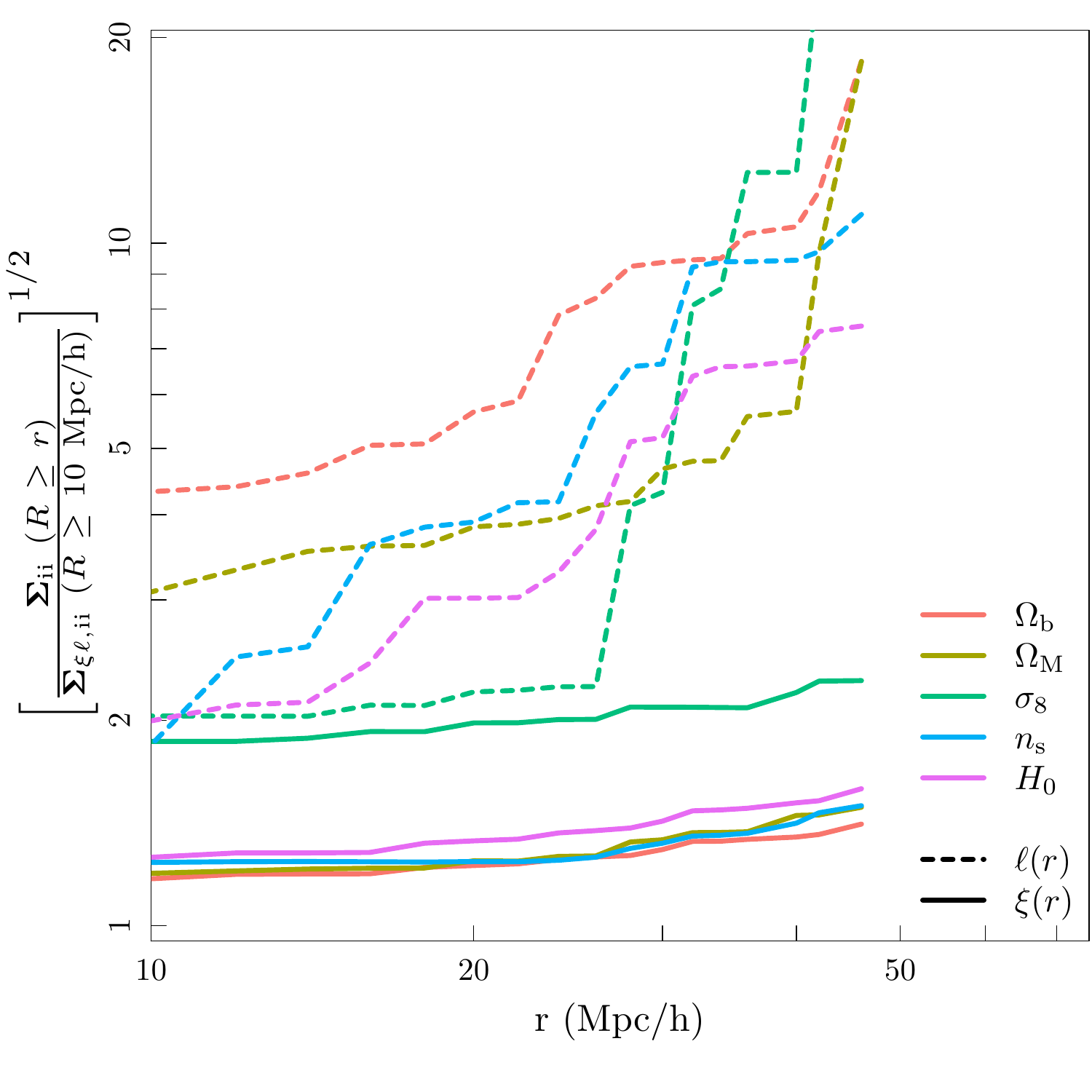}
	&
	\includegraphics[scale=0.525]{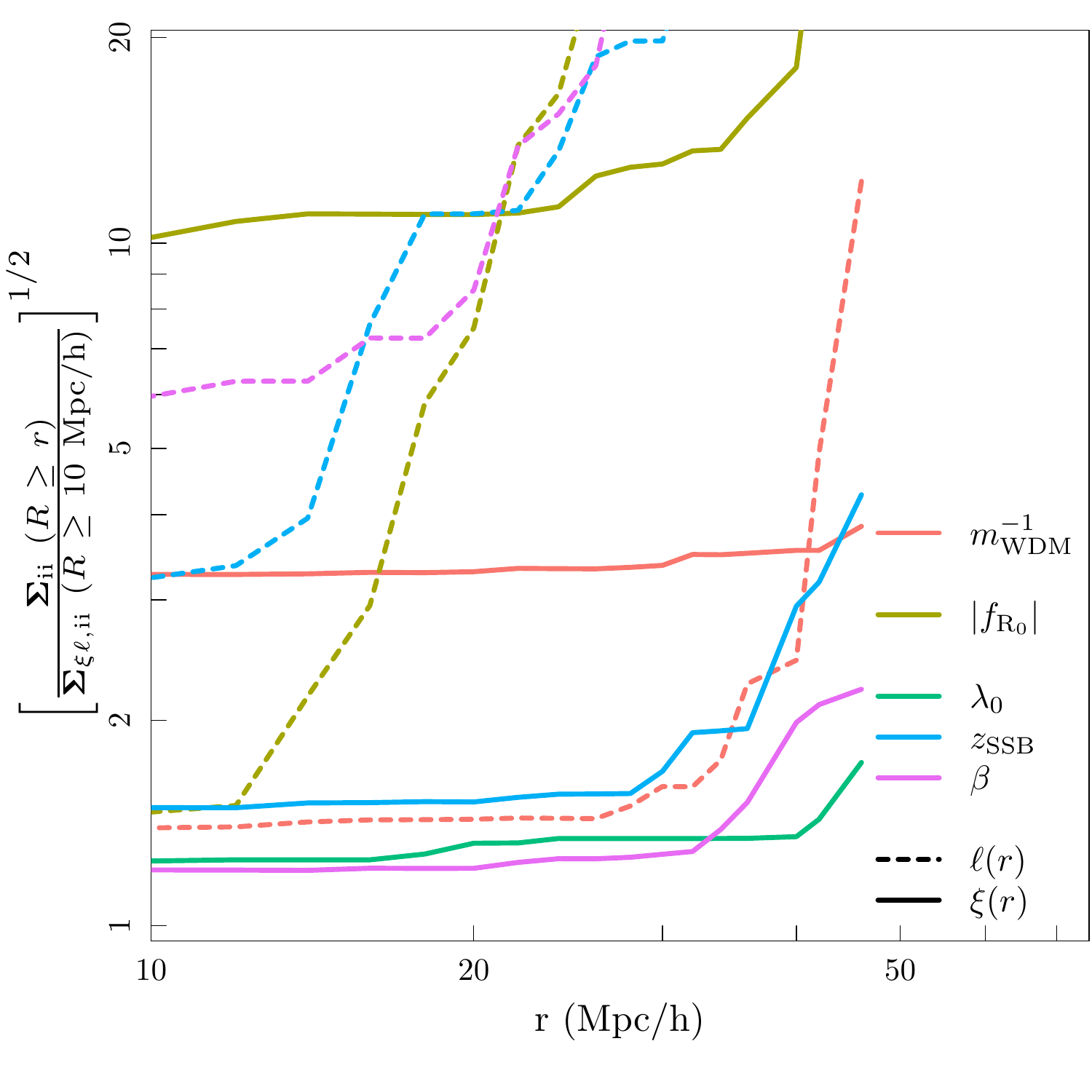}
\end{tabular}
    \caption{ The relative constraints placed on each standard parameter (left) and alternate cosmological parameter (right) as smaller scales are included in the FIM computation. The largest scales ($\geq 50 h^{3} {\rm Mpc}^{-3}$) suffer from statistical noise due to lack of modes and, hence, have been omitted from the figure. We find a sharp increase in constraining power of the LCF when compared to the 2PCF across all parameters. For this analysis we assume a fiducial cosmological survey with $\sigma_{\rm b}=0.13$, $\bar{n}=2\times 10^{-3} h^{3} {\rm Gpc}^{-3}$ and $\V=1 h^{-3} {\rm Gpc}^3$.}
    \label{fig:cosmo_info_scale}
\end{figure*}

\begin{figure*}
\centering
     \begin{center}
     \begin{tabular}{p{2cm} c}
     \toprule
       Cosmology & Parameter \\  \bottomrule
      \toprule\\
      \vspace{-2cm}  
       \begin{tabular}{c}
       WDM
       \end{tabular}
	&
	
       \includegraphics[trim={0 0 0 12.259cm},clip,scale=1]{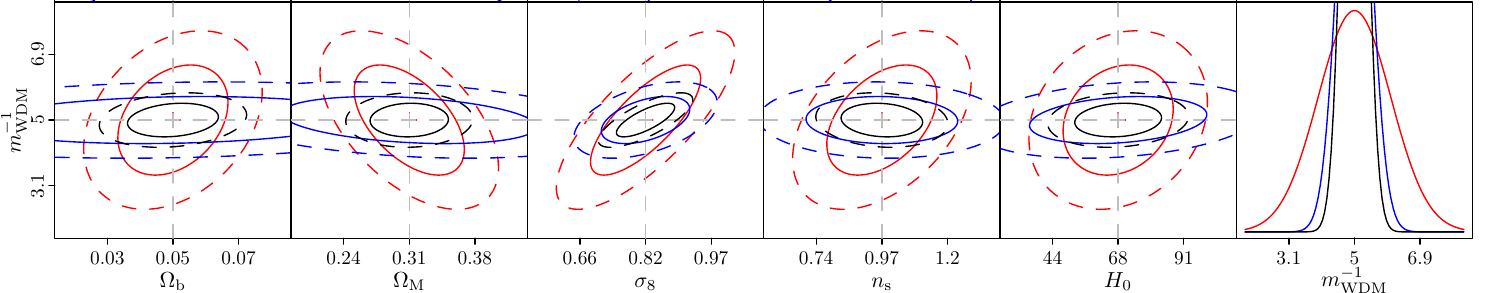}
      
      \\ \bottomrule \\
      \vspace{-2cm}  
       \begin{tabular}{c}
       f(R)
       \end{tabular}
       &

      \includegraphics[trim={0 0 0 12.259cm},clip,scale=1]{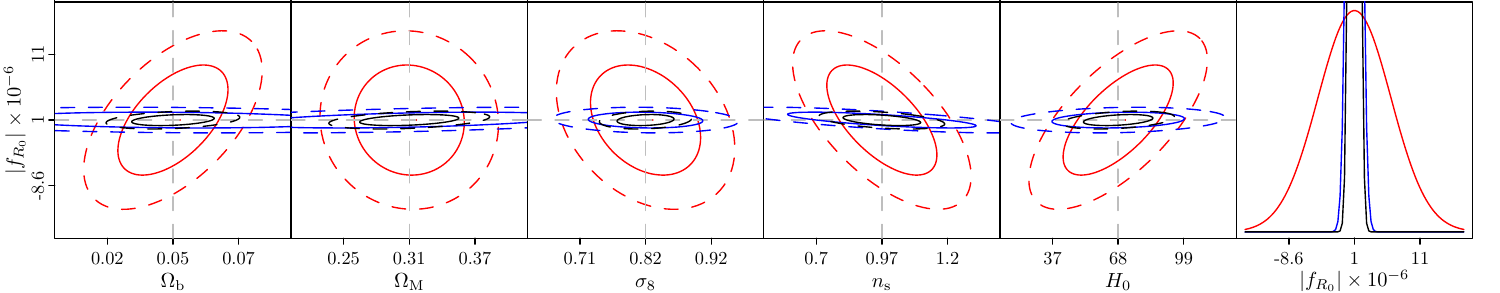}
      
      \\ \bottomrule \\
      
       \vspace{-3.3cm}  
       \begin{tabular}{c}
       Symmetron
       \end{tabular}
       &
      \includegraphics[trim={0 0 0 9.269cm},clip,scale=0.9]{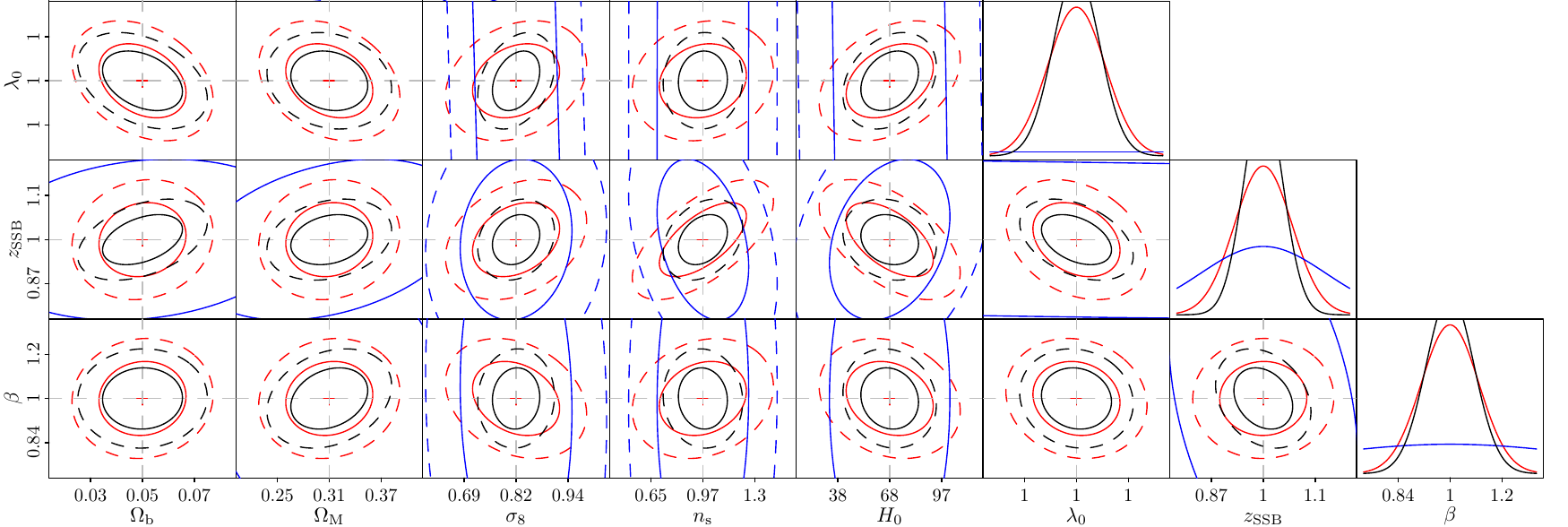}
      
      \\ \bottomrule
      
      \end{tabular}
      \caption{The 2PCF (red), the LCF (blue) and the combined (black) constraints for each of the alternate cosmological parameters determined by our analysis. The color scheme is same as that used in Fig.~\ref{fig:lcdm_fi}. The top, middle and bottom rows show the allowable range of WDM, $f(R)$ and Symmetron parameters, respectively, along with their pairwise constraints with the standard parameters. The elliptical orientations of the standard-alternate parameter pairs are significantly different for the 2PCF and LCF which allows a higher gain in constraining power for these models. The strongest constraints are comparable to that of $\sigma_{\rm 8}$ which is the best determined standard parameter following the FI analysis. The analysis assumes a fiducial survey with $13\%$ bias uncertainty, $2\times 10^{-3} h^{3} {\rm Gpc}^{-3}$ particle number density and $\V=1 h^{-3} {\rm Gpc}^3$.}
      \label{tbl:alternate_fim}
      \end{center}
      \end{figure*}

Fig.~\ref{tbl:alternate_fim} (top) shows the diagonal and off-diagonal components of $\Sigma$ determined by inverting the FIM. We find a \emph{gain} of 3.28 on the $m_{\rm WDM}$ parameter  with a non-noticeable change in constraints on the $\Lambda$CDM parameters. This is likely the result of the elliptical orientations of the 2PCF and the LCF being vastly different for $m^{-1}_{\rm WDM}$ and $\Lambda$CDM parameter pairs. Furthermore, the relative constraining power of the LCF for the $m^{-1}_{\rm WDM}$ increases faster when compared to that of standard parameters as shown in Fig.~\ref{fig:cosmo_info_scale} (right). Since the LCF is less sensitive to the linear growth its variation is mostly dictated by the local gravitational interactions and hence the properties of the underlying dark matter field. This attribute of the LCF allows it to contain significantly more information about the $m^{-1}_{\rm WDM}$ for the fiducial survey we assume in our analysis.

\subsection{f(R) Cosmology}

The first modified gravity (MG) analysed in this paper is a specific realisation of the Hu-Sawicki $f(R)$ model \citep{Hu2007PhRvD..76f4004H}. The original model was defined as a modification to the Ricci scalar, $R$, of the Einstein-Hilbert term, i.e. $R \rightarrow R+f(R)$, with a free function $f$ such that the action, $S$, for this model is given by
\begin{equation}
	S = \int \sqrt{-g}\left[\frac{R+f(R)}{16\pi G} + \mathcal{L}_m \right] d^4x,
\end{equation}
where $\mathcal{L}_m$ and $g$ are the matter Lagrangian and the Einstein frame metric, respectively. In this prescription the free function $f$ is defined as
\begin{equation}
	f(R) = -m^2\ \frac{c_{1} (R/m^2)^n}{c_{2} (R/m^2)^n + 1},
\end{equation}
where $m^2=H_0^2\Omega_{\rm M}$. The variables $c_1$, $c_2$ and $n$ are the free parameters for this MG model.

The ratio $c_1/c_2$ describes the expansion of the universe, and enforcing a $\Lambda$CDM-like expansion history reduces the number of free parameters to two, i.e. $n$ and $c_1$ (or $c_2$). In this case, the derivative $f_{\rm R} = df(R)/dR$, at present day ($z=0$), becomes
\begin{equation}
	f_{\rm R_0} = -n\ \frac{c_{1} }{c_{2}^2 } \left( \frac{\Omega_{\rm M}}{3 \left(\Omega_{\rm M} + 4\Omega_\Lambda\right)} \right)^{n+1}.
\end{equation}

The geodesic equation of this model takes the form
\begin{equation}
	\pmb{\rm \ddot{x}} + 2H\pmb{\rm \dot{x}} + \frac{\nabla\Phi}{a^2} - \frac{1}{2}\frac{\nabla f_{\rm R}}{a^2} = 0,
\end{equation}
where $a$ and $\Phi$ are the scale factor and the scalar perturbation (or the gravitational potential in the classical sense), respectively. The final term on left hand side is the additional `fifth' force beyond the standard gravity. The range of this force depends on the two free parameters and at $z=0$ can be quantified as
\begin{equation}
	\lambda_\phi^0 = 3\ \sqrt{\frac{n+1}{\Omega_{\rm M} + 4\Omega_\Lambda}}\ \sqrt{\frac{|f_{\rm R_0}|}{10^{-6}}} \ h^{-1}\rm Mpc,
\end{equation}
which is usually expressed in units of $h^{-1} \rm Mpc$. As mentioned earlier we only look at a special case of the Hu-Sawicki model where $n=1$. This allows us to fully define the the $f(R)$ model using a single parameter, $|f_{\rm R_0}|$. See \cite{Llinares2014A&A...562A..78L} for further details about this model and the reason for choosing this particular parameterisation.

The numerical simulations analysed in this paper are taken from \citep{Llinares2014A&A...562A..78L} which uses \texttt{Isis}, a derivative of the adaptive mesh refinement (AMR) code \texttt{RAMSES} \citep{Teyssier2002A&A...385..337T}, to fully evolve density fields sampled from an initial power spectrum generated by \texttt{LINGER} \citep{Bode1995astro.ph..4040B}. We evaluate the derivatives using the \texttt{fofr6} and \texttt{$\Lambda$CDM} runs under the assumption that the central cosmology is given by $|f_{\rm R_0}|=10^{-6}$ with the same spacing, i.e.~$\Delta|f_{\rm R_0}|=10^{-6}$. Since the standard $\Lambda$CDM and the $f(R)$ runs are both computed by the same codes with the same random seeds, the deviation between them purely results from the fifth force, allowing a robust calculation of the estimator derivative with respect to $|f_{\rm R_0}|$. Note that the $\Lambda$CDM parameters of our previous runs (Sections \ref{subsec:lcdm} and \ref{subsec:wdm}) slightly differ from those in \citep{Llinares2014A&A...562A..78L}, but we assume that this does not impact the derivatives with respect to $|f_{\rm R_0}|$.
Another difference is that the $\Lambda$CDM and $f(R)$ simulations of \cite{Llinares2014A&A...562A..78L} use a higher spatial resolution ($512^3$ particles in a $L=256\ h^{-1}\rm Mpc$ box). This higher resolution is required to properly take into account non-linear effects associated with the screening mechanism. Despite this increase in resolution, we restrict the FI computations to correlation scales $r\geq10\ h^{-1}\rm Mpc$ to be consistent with the previous cosmological models.

For the fiducial survey volume of $\V = 1\ h^{-3}\rm Gpc^3$ and a linear bias uncertainty of 13\% (standard deviation), the \emph{gain} on $|f_{\rm R_0}|$ constraints from including the LCF is $g=9.99$ as shown in Fig.~\ref{tbl:alternate_fim} (middle). The orientations of the 2PCF and the LCF ellipses are different for all $|f_{\rm R_0}|$ and $\Lambda$CDM parameter pairs which leads to the extra constraining power of the combined estimators. For this MG model, we find a sharp increase in the information content of the LCF at small scales as compared to that of the 2PCF as shown in Fig.~\ref{fig:cosmo_info_scale} (right). Hence, the LCF is more susceptible to non-linear growth in $f(R)$ model and requires us to measure scale down to $10\ h^{-1} {\rm Mpc}$.

\subsection{Symmetron}

The final cosmological model analysed in this paper is the Symmetron model. This MG model, which was originally explored by \citet{Hinterbichler2010PhRvL.104w1301H}, uses a scalar field $\phi$ governed by a potential and s conformal factor equation,
\begin{equation}
	V(\phi) = -\frac{1}{2}\mu^2\phi^2+\frac{1}{4}\lambda\phi^4,
\end{equation}
\begin{equation}
	A(\phi) = 1+\frac{1}{2}\frac{\phi^2}{M^2},
\end{equation}
where $\mu$ and $M$ are mass scales with $\lambda$ being a positive dimensionless constant. These three parameters define the simplest Symmetron model.

In this model, the fluctuations of the scalar field couple to the matter in regions of low density $\rho\lesssim\bar{\rho}$, where $\bar{\rho}$ is the average matter density of the universe. These correspond to cosmic voids. However, in over-dense regions, the coupling becomes negligible -- an effect known as `screening'. The vacuum expectation value of the scalar $\phi$ is
\begin{equation}
	\phi_0 = \frac{\mu}{\sqrt{\lambda}}
\end{equation}
and it determines the coupling strength between the scalar field and matter. For numerical convenience we rewrite the equations in terms of a dimensionless scalar field $\chi=\phi/\phi_{0}$ using the following parameters
\begin{equation}
\begin{split}
	\lambda_{\rm 0} &=\frac{1}{\sqrt{2}\mu} \\
	\beta &= \frac{\phi_{0}M_{\rm pl}}{M^{2}} \\
	a_{\rm SSB} &= \frac{\rho_{0}}{\mu^{2}M^{2}}
\end{split}
\end{equation}
where $M_{\rm pl}$ and $\rho_{0}$ are the Planckian mass and the background density at $z=0$, respectively. The physical interpretation of the parameters $\lambda_{\rm 0}$, $\beta$ and $z_{\rm SSB}$ are the range of the scalar field, its coupling strength and the redshift $z_{\rm SSB}$ (or its associated scale factor $a_{\rm SSB}$) at which the symmetry is broken (screening) for a given background cosmology, respectively. Using these parameters, the geodesic equation with the additional fifth force term becomes
\begin{equation}
	\pmb{\rm \ddot{x}} + 2H\pmb{\rm \dot{x}} + \frac{\nabla\Phi}{a^2} - 6\frac{\Omega_{\rm M}H^{2}_{0}}{a^{2}}\frac{\beta^{2}\lambda_{0}^{2}}{a_{\rm SSB}^3}\chi\nabla\chi = 0
\end{equation}
See \cite{Llinares2014A&A...562A..78L} for full details of this model and the reason for choosing this parametrisation.

We assume the Symmetron parameters are centred at $\bm\theta_0 = \{ \lambda_{\rm 0} (h^{-1} {\rm Mpc}), z_{\rm SSB}, \beta \} = \{1,1, 1 \}$ and use the \texttt{$\Lambda$CDM} box in conjunction with \texttt{Symm A-C} runs from  \citet{Llinares2014A&A...562A..78L} to evaluate the derivatives. Fig.~\ref{tbl:alternate_fim} (bottom) shows the parameter constraints obtained by employing the two estimators in our analysis. The LCF performs remarkably well, about as well as the 2PCF in constraining the three parameters. The gain of 2PCF+LCF, relative to the 2PCF alone, is about
$\bm{g}=\{1.24,1.48,1.21\}$. As with the previous MG model we find the orientation of the Symmetron and $\Lambda$CDM pair ellipses having different orientations for the two estimators. This provides a boost in the combined constraints. 

Fig.~\ref{fig:cosmo_info_scale} (right) indicates the importance of incorporating smaller scales in our analysis. In the case of Symmetron model, the relative constraints from the LCF decrease faster than the 2PCF. Although, the individual constraining power of the 2PCF is better than that of the LCF, small scale measurements still provide better combined constraint than standard parameters with the exception of $\sigma_{8}$.

\subsection{Effect of the linear bias uncertainty}\label{ss:linbias}

\begin{figure}
	\includegraphics[width=\columnwidth]{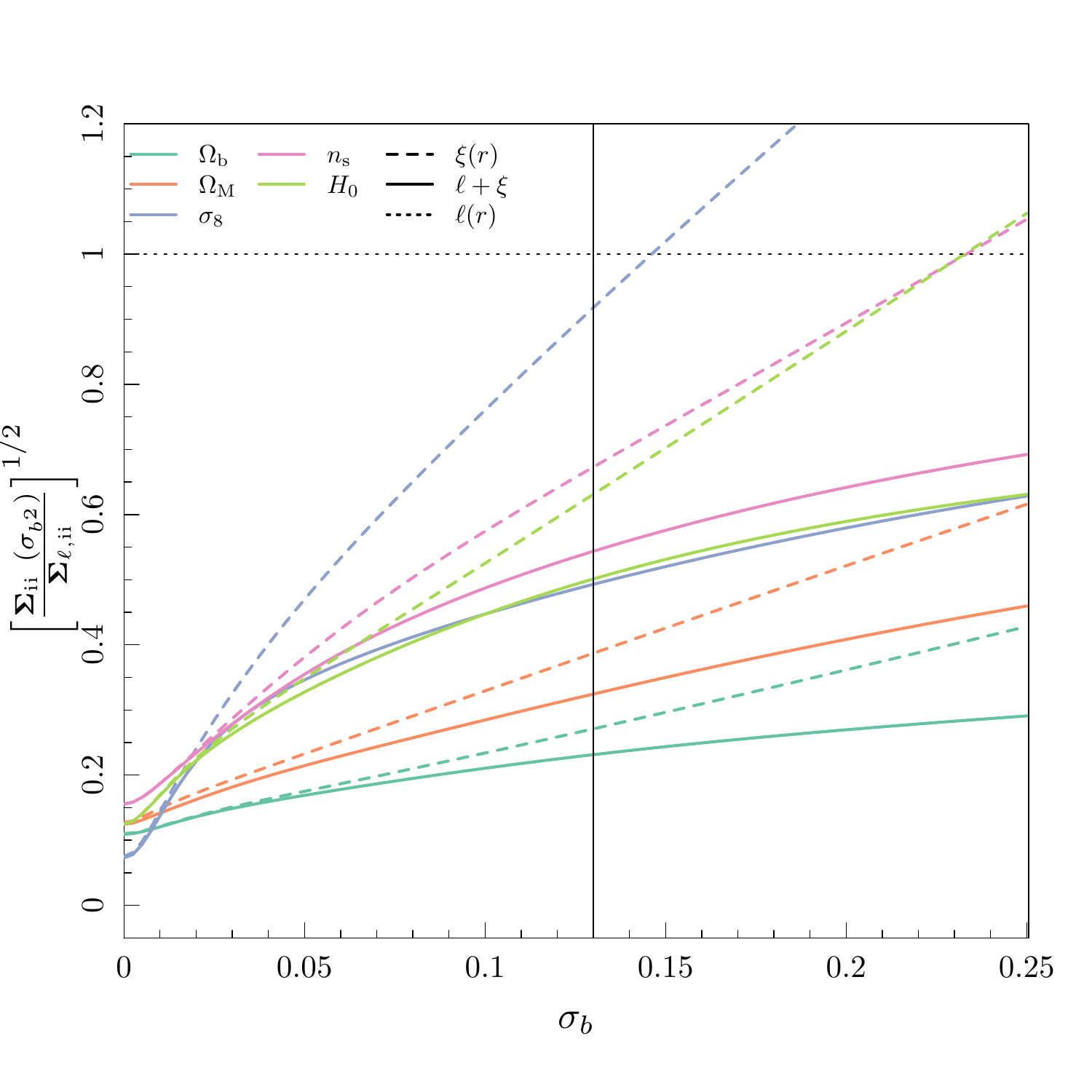} 

    \caption{The sensitivity of the $\Lambda$CDM constraints on the bias uncertainty, $\sigma_b$. The constraints placed by the LCF are independent of linear bias and, hence, have been used for normalisation. The solid vertical line shows the 13\% uncertainty used in this paper. As this parameter increases the constraints asymptote to that of the LCF.  The constraining power of the 2PCF pertaining to the $\sigma_{8}$ parameter has the sharpest decrease since this parameter is highly correlated with linear bias. The analysis assumes a fiducial survey volume of $\V = 1\ h^{-3}\rm Gpc^3$ and a space density of $\bar{n}=2\times 10^{-3} h^{3} {\rm Gpc}^{-3}$}
    \label{fig:bias_compare}
\end{figure}

\begin{figure}
	\includegraphics[width=\columnwidth]{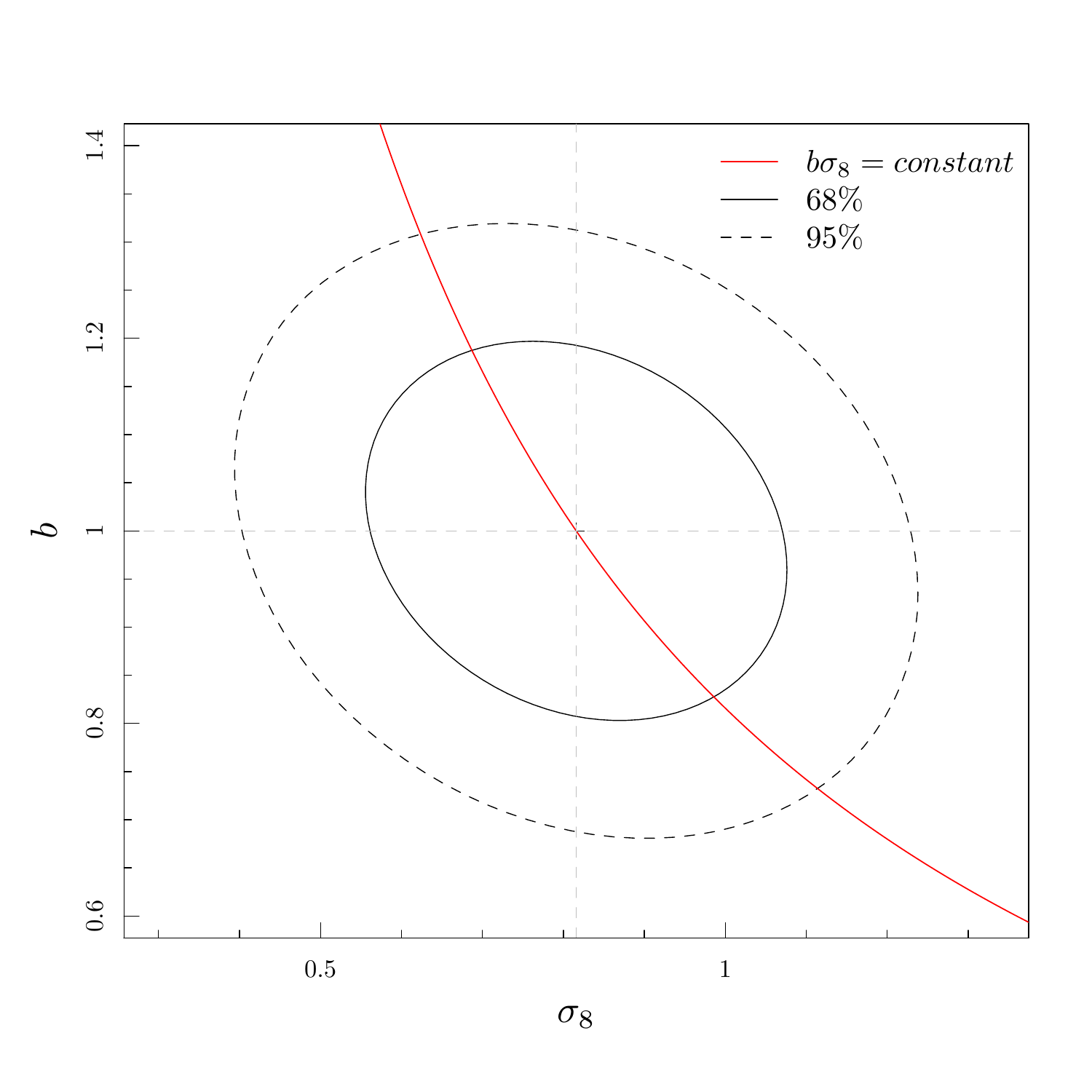} 

    \caption{The bounds placed on the $b$-$\sigma_{\rm 8}$ using the combined information from the 2PCF and the LCF. The ellipses show the 68\% (solid) and 95\% (dotted) Gaussian contours inferred by only using length scales $\geq 50 h^{-1}\rm Mpc$. For smaller scales the non-linear correction to bias becomes significant and hence the constraints become unreliable. If we were to carry out a full likelihood analysis we expect the constraints to be parallel to the $b$-$\sigma_{\rm 8}$ curve (solid red line). We assume a fiducial survey volume of $\V = 1\ h^{-3}\rm Gpc^3$ to scale the covariance matrix  and sub-sample particles to a final number density of $\bar{n}=2\times 10^{-3} h^{3} {\rm Gpc}^{-3}$.}
    \label{fig:bias_s8_ellipse}
\end{figure}

The statistical uncertainty of the galaxy (or halo) bias deteriorates the information extracted from cosmic large-scale structure. At linear order, this effect only applies to the 2PCF, not to the LCF, since the latter is insensitive to the linear bias $b$. Hence, the information of the LCF relative to the 2PCF increases with the uncertainty of $b$. So far, we have accounted for this effect assuming that $b$ has a fixed normal uncertainty of $\sigma_{\rm b}=13\%$. The variation of the relative FI with this uncertainty is shown in Fig.~\ref{fig:bias_compare}. Since the relative information in the 2PCF decreases with increasing $\sigma_{\rm b}$, the combined information asymptotes to that of the LCF estimator as $\sigma_b\rightarrow\infty$. In all cosmological models the LCF adds significant extra constraints to the 2PCF, if $\sigma_{\rm b}=0.13$ (as in Fig.~\ref{fig:lcdm_fi} and Fig.~\ref{tbl:alternate_fim}) or larger, whereas, in a scenario where bias is perfectly constrained, the 2PCF constrains each parameter by up to 4 times.

The discussion above highlights the power of combining the two-point statistics with a linear bias-independent estimator for parameter estimation. In turn, this property of the LCF also allows us to infer the bias parameter $b$ itself. This can be achieved by considering $b$ as an additional free parameter (see first approach discussed in Section \ref{subsec:bias_definition}) and computing the full FIM of all cosmological parameters and $b$. This requires the derivative of the galaxy 2PCF with respect to $b$, whose analytical expression is simply
\begin{equation}
	\frac{\partial\xi_{\rm g,i}}{\partial b}= 2b\xi_{\rm i}.
\end{equation}
Since we now treat $b$ as an additional model parameter, its uncertainty must not be included in the covariance via equation~(\ref{eq:diagonalscaling}).
Importantly, the linear bias $b$ is only a good approximation of the full bias on large scales ($r\geq 50 h^{-1}\rm Mpc$). Hence, we restrict this analysis to these large scales, and maintain the fiducial survey volume of $\V = 1\ h^{-3}\rm Gpc^3$. The LCF at these scales is susceptible to statistical noise, hence, we use 50 \texttt{L-PICOLA} to better estimate the large scale derivatives of the 2PCF and the LCF for each standard parameter. The resulting combined constraints on $\sigma_{\rm 8}$ and $b$ are shown in Fig.~\ref{fig:bias_s8_ellipse}. In the Laplace approximation (Gaussian likelihood) these constraints correspond to elliptical uncertainties. However, from the collapse theory of haloes, $\sigma_{8}b$ is constant if the other cosmological parameters are fixed. This means that the degeneracy between $\sigma_{8}$ and $b$ corresponds to a hyperbola shown as solid curve in Fig.~\ref{fig:bias_s8_ellipse}. As expected, the major axis of the ellipses roughly align with the direction of this hyperbola. Quantitatively, this analysis finds a linear bias uncertainty of $\sigma_{\rm b} \approx 13\%$ which we use throughout this study. We also compare the constraints found by the two methods for each of the $\Lambda$CDM parameters and find similar results with $\sim 10\%$ difference.
\section{Conclusions} \label{sec:conclusion}

In this paper we have combined two-point statistics and the line correlation function, three-point Fourier phase estimator, to infer cosmological parameters in standard and non-standard cosmological models. The Fisher information matrix was used to quantify the information in these estimators and derive parameter posteriors.

In the absence of linear bias uncertainties, the extra information provided by the LCF in addition to that already present in the 2PCF is marginal (about 5\% on average). However, for a fiducial linear bias uncertainty of $13\%$ and a survey volume of $\V = 1\ h^{-3}\rm Gpc^3$, we found that the addition of the LCF improves the 2PCF-based parameter constraints by a significant factor of about $\sim1.2$ (i.e.\  parameter uncertainties become 1.2-times smaller) in standard $\Lambda$CDM cosmology and by a factor up to $\sim 1.2-10$ in MG models and $\sim 3.3$ in WDM cosmology. The relative information in the LCF increases with increasing bias uncertainty. However, to fully gauge the usefulness of the LCF we need to take into account non-linear bias terms in the computation. The constraints determined by the LCF in this study are likely an over-estimate however our treatment simplifies the methodology and can be regarded as a first step towards more realistic constraints. To optimally benefit from the information in the LCF scales down to about $r=10\ h^{-1}\rm Mpc$ should be resolved. The combination of the 2PCF and the LCF can also be used to infer the linear bias, which is hard to measure otherwise and consequently to break the degeneracy between bias and $\sigma_8$.

Overall, these results advocate the use of Fourier phase statistics in addition to standard two-point statistics (2PCF of power spectrum) when inferring cosmological parameters from modern galaxy redshift surveys. The Fourier phase space is an excellent probe of the local interactions between the dark matter particles and even at larger scales before the full onset of non-linear gravitational growth the LCF provides more information about the properties of the underlying dark matter density field than the 2PCF. The gain from including the LCF is especially important in modified theories of gravity like in the $f(R)$ and Symmetron models. This reflects the fact that modified gravity models possess a very rich phenomenology in the non-linear regime, or more precisely at the transition from the linear to the non-linear regime, where the screening mechanism takes place. As a consequence the LCF which probes directly the emergence of non-linear correlations is very sensitive to these modifications of gravity and provides stronger constraints than the 2PCF.

The analysis of phase statistics comes with its own caveats. Apart from being computationally quite expensive, phase statistics such as the LCF are still difficult to measure in real surveys with irregularly shaped non-periodic volumes and complex selection functions. Clever methods for extracting such phase statistics, similarly to those used for the 2PCF \citep{LS1993}, remain yet to be developed. Furthermore, redshift space distortions have their own perturbing effects \citep{Eggemeier2015MNRAS.453..797E}, which, when harvested carefully, might be used for additional cosmological constraints.

\section*{Acknowledgements}

We thank our referee for helpful comments to improve the paper. This work was supported by the Research Collaboration Award 12105205 of the University of Western Australia. D.O. acknowledges support from Australian Research Council grants DP160102235. CLL acknowledges support from STFC consolidated grant ST/L00075X/1 \& ST/P000541/1. This work used the DiRAC Data Centric system at Durham University.  This equipment was funded by BIS National E-infrastructure capital grant ST/K00042X/1, STFC capital grants ST/H008519/1 and ST/K00087X/1, STFC DiRAC Operations grant ST/K003267/1 and Durham University. DiRAC is part of the National E-Infrastructure. Parts of this research were conducted by the Australian Research Council Centre of Excellence for All-sky Astrophysics (CAASTRO), through project number CE110001020. Part of the simulations were done on the Sciama High Performance Compute (HPC) cluster which is supported by the ICG, SEPNet and the University of Portsmouth. CB 
and FOF acknowledge support by the Swiss National Science Foundation.




\bibliographystyle{mnras}
\bibliography{draft.bib}








\bsp	
\label{lastpage}

\end{document}